\documentclass[prl,twocolumn,amsmath,amsfonts,amssymb,showpacs]{revtex4} 
\usepackage{epsfig,psfrag,graphicx,bm,color}

\def\del#1{{}}

\newcommand{\beq}{\begin{equation}}
\newcommand{\eeq}{\end{equation}}

\newcommand{\dd}{\mathrm{d}}
\newcommand{\mx}{\ensuremath{m_{\chi}}}
\newcommand{\ngamma}{\ensuremath{N_{\gamma}}}
\newcommand{\ngammai}{\ensuremath{N_{\gamma,i}}}
\newcommand{\sigmaannv}{\ensuremath{\langle\sigma v\rangle}}
\newcommand{\sigv}{\ensuremath{\sigma_v}}

\newcommand{\egamma}{\ensuremath{E_{\gamma}}}
\newcommand{\fsr}{\mathrm{fsr}}

\newcommand{\rs}{\ensuremath{r_s}}

\newcommand{\rmn}{\mathrm}
\newcommand{\e}{\rmn{e}}
\newcommand{\B}{\rmn{B}}

\begin{document}

\title{Gamma-rays from dark matter annihilations strongly constrain the
  substructure in halos}

\author{Anders Pinzke$^{1}$} \email{apinzke@fysik.su.se} 
\author{Christoph Pfrommer$^{2}$}\email{pfrommer@cita.utoronto.ca}
\author{Lars Bergstr\"om$^{1}$}\email{lbe@fysik.su.se}

\affiliation{$^{1}$The Oskar Klein Centre for Cosmoparticle Physics, Department
  of Physics, Stockholm University, AlbaNova University Center, SE - 106 91
  Stockholm, Sweden}

\affiliation{$^{2}$Canadian Institute for Theoretical Astrophysics,
  University of Toronto, 60 St. George Street, Toronto, Ontario, M5S
  3H8, Canada}

\date{\today}


\begin{abstract}
 Recently, it has been shown that electrons and positrons from dark matter (DM)
 annihilations provide an excellent fit to the Fermi, PAMELA, and HESS data.
 Using this DM model, which requires an enhancement of the annihilation cross
 section over its standard value to match the observations, we show that it
 immediately implies an observable level of $\gamma$-ray emission for the Fermi
 telescope from nearby galaxy clusters such as Virgo and Fornax. We show that
 this DM model implies a peculiar feature from final state radiation that is a
 distinctive signature of DM.  Using the EGRET upper limit on the $\gamma$-ray
 emission from Virgo, we constrain the minimum mass of substructures within DM
 halos to be $ > 5\times10^{-3}\,\rmn{M}_\odot$ -- about four orders of
 magnitudes larger than the expectation for cold dark matter. This limits the
 cutoff scale in the linear matter power spectrum to $k < 35\, \rmn{kpc}^{-1}$
 which can be explained by e.g., warm dark matter. Very near future Fermi
 observations will strongly constrain the minimum mass to be $ >
 10^{3}\,\rmn{M}_\odot$: if the true substructure cutoff is much smaller than
 this, the DM interpretation of the Fermi/PAMELA/HESS data must be wrong.  To
 address the problem of astrophysical foregrounds, we performed high-resolution,
 cosmological simulations of galaxy clusters that include realistic cosmic ray
 (CR) physics. We compute the dominating $\gamma$-ray emission signal resulting
 from hadronic CR interactions and find that it follows a universal spectrum and
 spatial distribution. If we neglect the anomalous enhancement factor and assume
 standard values for the cross section and minimum subhalo mass, the same model
 of DM predicts comparable levels of the $\gamma$-ray emission from DM
 annihilations and CR interactions. This suggests that spectral subtraction
 techniques could be applied to detect the annihilation signal.
\end{abstract}

\maketitle

The identity of the DM in the Universe has been the subject of intense
speculation. In particular, the hierarchical formation of structure, as
indicated from numerical simulations of Cold DM (CDM), agrees very well with
observations on scales of galaxy clusters and larger, whereas the small scale
behaviour on galactic and subgalactic scales is more unsecure. If dark matter
particles have weak interactions, one would expect possible signals from
annihilations (or decays).

Data from a new generation of cosmic ray detectors have indeed been tentatively
interpreted in terms of such signatures of DM. In particular, the positron
fraction measured by the PAMELA satellite \cite{pamela_positrons} and the sum of
electrons and positrons by ATIC \cite{atic} have shown an unexpected excess.
Very recent data from Fermi-LAT \cite{fermi,fermicre} and H.E.S.S. \cite{hess}
on the sum of electrons and positrons do not confirm the peak claimed by ATIC,
but still indicate an excess compared to the expected background in conventional
models.  A number of attempts have been made trying to explain the excess due to
DM annihilation (for recent reviews, see \cite{hooper,lberev}) but also other
astrophysical sources such as pulsars have been investigated (e.g.,
\cite{pulsars}, and references therein).

One class of DM models that fits the new data has halo annihilation primarily
into muon pairs, which then decay to electrons and positrons. In \cite{bez}
examples of fits with remarkable quality (which also fit PAMELA
\cite{pamela_positrons} and new H.E.S.S.  data) were obtained by such DM
models. It was pointed out that if the annihilation goes directly into a
$\mu^+\mu^-$ pair, a striking signature may be present. This is caused by the
direct emission of $\gamma$-rays from the final state (final state radiation,
FSR), which gives a peculiar energy spectrum, with $E^2dN_\gamma/dE$ almost
linearly rising with energy. The same, but weaker, feature may in fact exist
also for the theoretically perhaps more easily motivated models with
intermediate spin-0 boson decay, but in this Letter we only treat the somewhat
simpler direct annihilation case. In this Letter we show that such DM models
face considerable tension with existing $\gamma$-ray limits from clusters of
galaxies, systems which will be very interesting to detect and study with coming
$\gamma$-ray detectors (Fermi, H.E.S.S., MAGIC, VERITAS, and eventually CTA).

Galaxy clusters constitute the most massive objects in our Universe that are
forming today. This causes their DM subhalo mass function to be less affected by
tidal stripping compared to galaxy sized halos that formed long ago.  The
annihilation luminosity of the smooth DM halo component scales as
\begin{equation}
\label{eq:DMlum}
  L_\rmn{sm} \sim \int \dd V \rho^2 \sim \frac{M_{200}\,
    c^3}{[\log(1+c)-c/(1+c)]^2} \sim M_{200}^{0.83},
\end{equation}
where the virial mass $M_{200}$ and the concentration $c$ (see
Eqn.~\ref{eq:cfit}) are the two characteristic parameters of the universal
Navarro-Frenk-White (NFW) density profile $\rho_\rmn{NFW}$ of DM halos
\cite{NFW}.  Hence, the flux ratio of a nearby cluster (Virgo) to a prominent
dwarf spheroidal (Draco) is given by
\begin{equation}
  \label{eq:orderofmag}
  \frac{F_\rmn{cluster}}{F_\rmn{dwarf}} \simeq
  \left(\frac{80\,\rmn{kpc}}{17\,\rmn{Mpc}}\right)^2
  \left(\frac{2\times10^{14}\,\rmn{M}_\odot}{10^{8}\,\rmn{M}_\odot}\right)^{0.83}
  \simeq 3.8,
\end{equation}
assuming an early formation epoch of the dwarf galaxy before the end of
reionization \cite{Draco}.  Once a satellite galaxy is accreted by our Galaxy,
the outer regions are severely affected by tidal stripping. The longer a
satellite has been part of our Galaxy, and the closer it comes to the center
during its pericentral passage, the more material is removed \cite{subhalo}. In
contrast, the substructure in clusters is not affected in the outer regions and
may enhance the DM annihilation signal over its smooth contribution considerably
as we will see in the following.  The FSR feature of DM annihilation may in
addition be more easily visible in clusters, as the average intensity of
starlight, which may give a masking signal due to inverse Compton scattering of
the copiously produced electrons and positrons, is lower than in the Milky Way
MW.  For previous work related to dark matter in clusters, see, e.g.,
\cite{colafrancesco, previous}. All halo masses and length scales are scaled to
the currently favored value of Hubble's constant, $H_0 = 70\,
\rmn{km~s}^{-1}\,\rmn{Mpc}^{-1}$. We define the virial mass $M_{200}$ and virial
radius $r_{200}$ as the mass and radius of a sphere enclosing a mean density
that is 200 times the critical density of the Universe $\rho_\rmn{cr}$.

{\bf Method.}  As our default model for {\em dark matter annihilation}, we take
the Sommerfeld-enhanced (see, e.g., \cite{sommerfeld, resonance}) direct muon
annihilation mode of \cite{bez}, i.e., mass $m_\chi=1600$ GeV and effective
enhancement factor 1100 relative to the standard annihilation cross section
$\sigmaannv_0 \sim 3 \times 10^{-26} \,\rmn{cm}^{3}\, \rmn{s}^{-1}$. It is
non-trivial to rescale this boost to the corresponding value for the cluster
environment. This may either give a smaller or larger value, depending, e.g., on
the velocity dispersion of bound substructure in the cluster, and whether the
Sommerfeld enhancement (SFE) increases down to very small velocities, or if it
saturates at some minimum velocity \cite{SFEformula}. We choose a simple and
generic model for the SFE factor $\mathrm{B_{sfe}}\left(\sigv\right)\thickapprox
0.7\, c/\sigma_v$ and saturate at $\sigma_{v,\,\rmn{min}} = 200
\,\rmn{km/s}$ \footnote{This velocity scale is a conservative choice for our DM
  model under consideration. We note that local anisotropy in the dark matter
  distribution, resonance effects of the cross section, and velocity
  uncertainties at freeze-out bring the best-fit enhancement factor of 1100
  \cite{resonance} easily into agreement with conventional models.}. For a given
cluster, we assume a constant velocity dispersion of $\sigma_v = 960
\,\rmn{km/s}\times ( M_{200}/ 10^{15}\,M_\odot)^{1/3}$ \cite{Voit}. This scaling
would result in a boost factor of $\mathrm{B_{sfe}} = 220$ for a cluster with
$M_{200} = 10^{15}\,M_\odot$.

Using cluster masses from the complete sample of the X-ray brightest clusters
(the extended HIFLUGCS catalogue, \cite{HIFLUGCS}) we identify the brightest
clusters for DM annihilation. In models with SFE, the DM flux to leading order
scales as a power-law $F \sim M_{200}^{-1/3} M_{200}^{0.83\phantom{/}}/D^2$. The
first factor accounts for the SFE and the second one is derived from
Eqn.~\ref{eq:DMlum}, using a power-law fit to the mass dependence of the NFW
halo concentration derived from cosmological simulations with $M_{200}
\gtrsim 10^{10} \rmn{M}_\odot$ \cite{maccio_cfit},
\begin{equation}
\label{eq:cfit}
  c=3.56 \times \left(\frac{M_{200}}{10^{15}\,\rmn{M}_\odot}\right)^{-0.098}\, .
\end{equation}
Note that Eqn.~\ref{eq:cfit} agrees well with \cite{zhao_cfit} for cluster-mass
halos after converting the concentration definitions according to \cite{c_conv}.
This yields Fornax ($M_{200}=10^{14}\,M_\odot$) and Virgo
($M_{200}=2.1\times10^{14}\,M_\odot$, \cite{M_Virgo}) as the prime targets for
DM observations and we additionally decide in favor of the well studied cluster
Coma ($M_{200}=1.4\times10^{15}\,M_\odot$) for comparison.

The differential photon flux from DM annihilation within a given solid
angle $\Delta \Omega$ along a line-of-sight (los) is given by
\begin{equation}
\label{eq:dflux}
\frac{\dd F}{\dd E_\gamma} \equiv
\frac{\dd^3 \ngamma}{\dd A \,\dd t\, \dd \egamma}
= \int_{\Delta\Omega}
\frac{\dd\Omega}{4\pi} \int_{\rm los} \dd l\, q_{\rm
  sm}\left(\egamma,r\right)\B_\rmn{F}(\sigv,r),
\end{equation}
where $q_{\rm sm}\left(\egamma,r\right)$ is the source function from the smooth
halo with contributions from two main processes: DM annihilating to
$\mu^+/\mu^-$ which decay to $\rm{e}^+/\rm{e}^-$ pairs that Compton up-scatter
CMB photons (IC) and FSR.  The source function of FSR is given by
\begin{equation}
q_\fsr \left(\egamma,r\right) = \sum_i \frac{\dd
  \ngammai}{\dd E_\gamma} \Gamma_i(r) ,
\end{equation}
where the annihilation rate density $\Gamma_i = (\rho/\mx)^2 \,
\sigmaannv_i/2$. The $i$ runs over all $\gamma$-ray producing channels each with
the spectrum $\frac{\dd \ngammai}{\dd E_\gamma}$ and annihilation cross-section
$\sigmaannv_i$. We use the standard photon distribution for the final state
radiation from our DM model annihilating directly to charged leptons assuming
$m_\chi \gg m_l$ \cite{Cholis:2008wq}. For the remaining part of this work,
the Einasto density profile for DM halos \cite{Einasto} is used, normalized with
$\rho_0 = \rho_\mathrm{NFW}(\rs)/4$ relying on the assumption that $90 \%$ of
the flux from a NFW density profile and an Einasto density profile originate
from within the scale radius $\rs = r_{200}/c$.

The product of enhancement factors from SFE $\B_\rmn{sfe}(\sigv)$ and from
substructure enhancement over the smooth halo contribution
$\B_\rmn{sub}\left(r\right) = 1+q_\rmn{sub}(r)/q_\rmn{sm}(r)$ is denoted by
$\B_\rmn{F}(\sigv,r) = \B_\rmn{sfe} (\sigv)\,\B_\rmn{sub}(r)$.  High-resolution
DM simulations of the MW suggest an enhancement from substructures
of approximately 220 inside $r_{200}$ assuming that the subhalos
  extrapolate smoothly down from the simulation resolution limit to smallest
  scales \cite{Aquarius}, with most of the substructure residing in the outer
part of the MW halo. We fit the luminosity $\rmn{L_{sub}}= \int \dd E_\gamma\,
\dd V q_\rmn{sub}$ from substructures inside a radius $r$ following
\cite{subboost},
\begin{equation}
  \mathrm{L_{sub}}\left(<r\right)=0.8 C\,{\rm L_{sm}}(r_{200})\,
  \left(r/r_{200}\right)^{0.8 \left(r/r_{200}\right)^{-0.315}},
  \label{eq:LSub}
\end{equation}
where ${\rm L_{sm}}(r_{200})$ is the smooth cluster halo luminosity within
$r_{200}$. The normalization $C=(M_{\rm min}/M_{\rm lim})^{0.226}$, where
$M_{\rm min}=10^5 M_\odot$ is the minimum subhalo mass in the simulation and
$M_{\rm lim}$ the free streaming mass. While its conventional value is $10^{-6}
M_\odot$ \cite{Mlim}, we will constrain this quantity by requiring consistency
with the non-detection of $\gamma$-ray emission from clusters by EGRET: the
smaller $M_{\rm lim}$, the more substructure is present and the larger is the
expected $\gamma$-ray signal. This approach of fitting the scaling behaviour of
$\mathrm{L_{sub}}(M_\rmn{lim})$ directly from numerical simulations
self-consistently accounts for the radial dependence of the substructure
concentration \cite{Aquarius}. We note that this might result in a slight
overestimate of the substructure luminosity if the assumed power-law scaling
flattens towards smaller scales although current simulations show no sign of
such a behavior which is also not expected since we are approaching the
asymptotic behavior in the power spectrum on these scales.

The source function of inverse Compton emission resulting from DM annihilating
is given by
\begin{equation}
  q_{\rm IC}\left(E_\gamma, r\right) = \int
  \dd E_\e\, \frac{\dd n_\e}{\dd E_\e} 
  P_{\rm IC}\left(E_\gamma,E_\e\right) ,
  \label{eq:ICemiss}
\end{equation}
where $P_{\rm IC}$ is derived by convolving the IC cross-section with the
differential target photon number density \cite{colafrancesco}.  Assuming that
the spatial diffusion time scale is much larger than the energy loss time scale,
the total equilibrium distribution of the electrons plus positrons is given by
\begin{eqnarray}
\left({\frac{\dd n_\e}{\dd E_\e}}\right)\left(E_\e, r \right) &=&
 \frac{\Gamma_\mu(r)}{b(E_\e, r)} \int_{E_\e}^{\mx c^2} \dd E_\e' \, 
  \frac{\dd N_\e}{\dd E_\e'},
\label{eq:nds}\\
b(E_\e, r) &=& \frac{4\sigma_{\rm T}c}{3(m_{\rm e}c^2)^2}\frac{B^2_{\rm
    CMB}+B^2(r)}{8\pi} E_\e^2,
\end{eqnarray}
where $\frac{\dd N_\e}{\dd E_\e}$ denote the differential number of electrons
plus positrons resulting from an annihilation event, $B_{\rm CMB}=3.24(1+z)^2\mu
{\rm G}$ denotes the equivalent field strength of the CMB, and we parametrize
the magnetic field of the galaxy cluster by $B(r) = 3\mu\rmn{G} \,[n_{\rm
    e}(r)/n_{\rm e}(0)]^{0.7}$, which follows from flux frozen magnetic fields.

To address the problem of source confusion by {\em astrophysical backgrounds},
we perform high-resolution, cosmological simulations \cite{Springel:2005mi} of a
sample of 14 galaxy clusters \cite{paperII}. They span over one and a half
decades in mass and follow radiative gas physics, star formation, supernova
feedback. In particular, we use an updated version of the cosmic ray physics
that is capable of following the spectral evolution of the cosmic ray (CR)
distribution function by tracking multiple CR populations -- each being
described by its characteristic power-law distribution with a distinctive slope
that is determined by the acceleration process \cite{TeVpap}.  We compute the
dominating $\gamma$-ray emission signal from decaying neutral pions that result
from hadronic CR interactions with the ambient gas following \cite{piondecay}.
We find that it obeys a universal spectrum and spatial distribution (details
will be presented in \cite{TeVpap}). This allows us to reliably model the cosmic
ray signal from nearby galaxy clusters using their true density profiles as
obtained by X-ray measurements \cite{electdens} that we map onto our simulated
density profiles. We compute $\gamma$-ray luminosity-mass scaling relations of
our sample \cite{paperIII} and use these to normalize the CR induced emission of
all clusters in HIFLUGCS \cite{HIFLUGCS}. In our optimistic CR model, we
calculate the cluster's total $\gamma$-ray flux within a given solid angle while
we cut the emission from our individual galaxies and compact galactic-sized
objects in our more conservative baseline model
\footnote{These dense gas clumps are being stripped off their galaxies due to
  ram pressure and dissociate incompletely in the intra-cluster medium due to
  insufficient numerical resolution as well as so far incompletely understood
  physical properties of the cluster plasma.}.

\begin{figure*}
\begin{minipage}{2.0\columnwidth}
  \includegraphics[width=0.48\columnwidth]{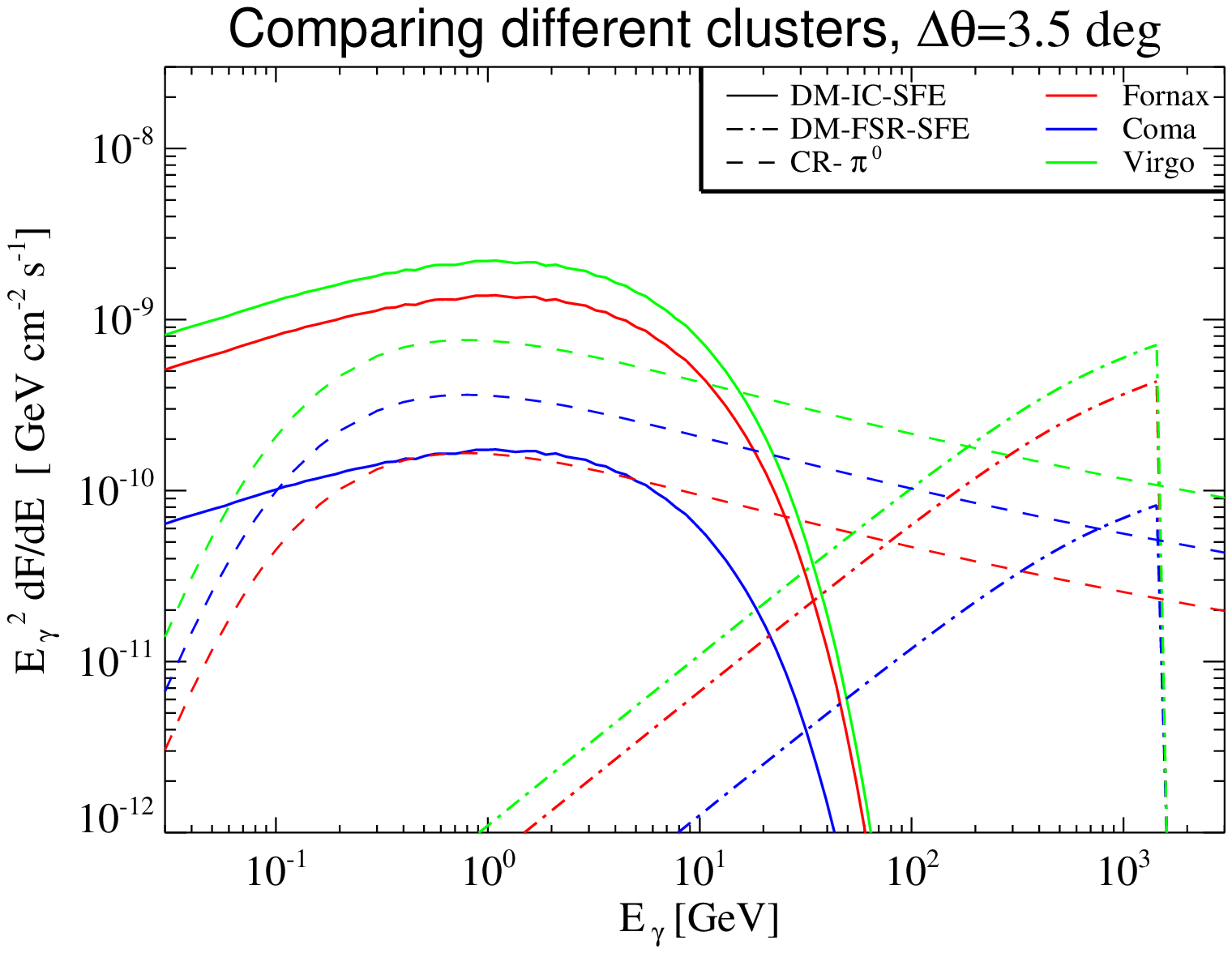}
  \includegraphics[width=0.48\columnwidth]{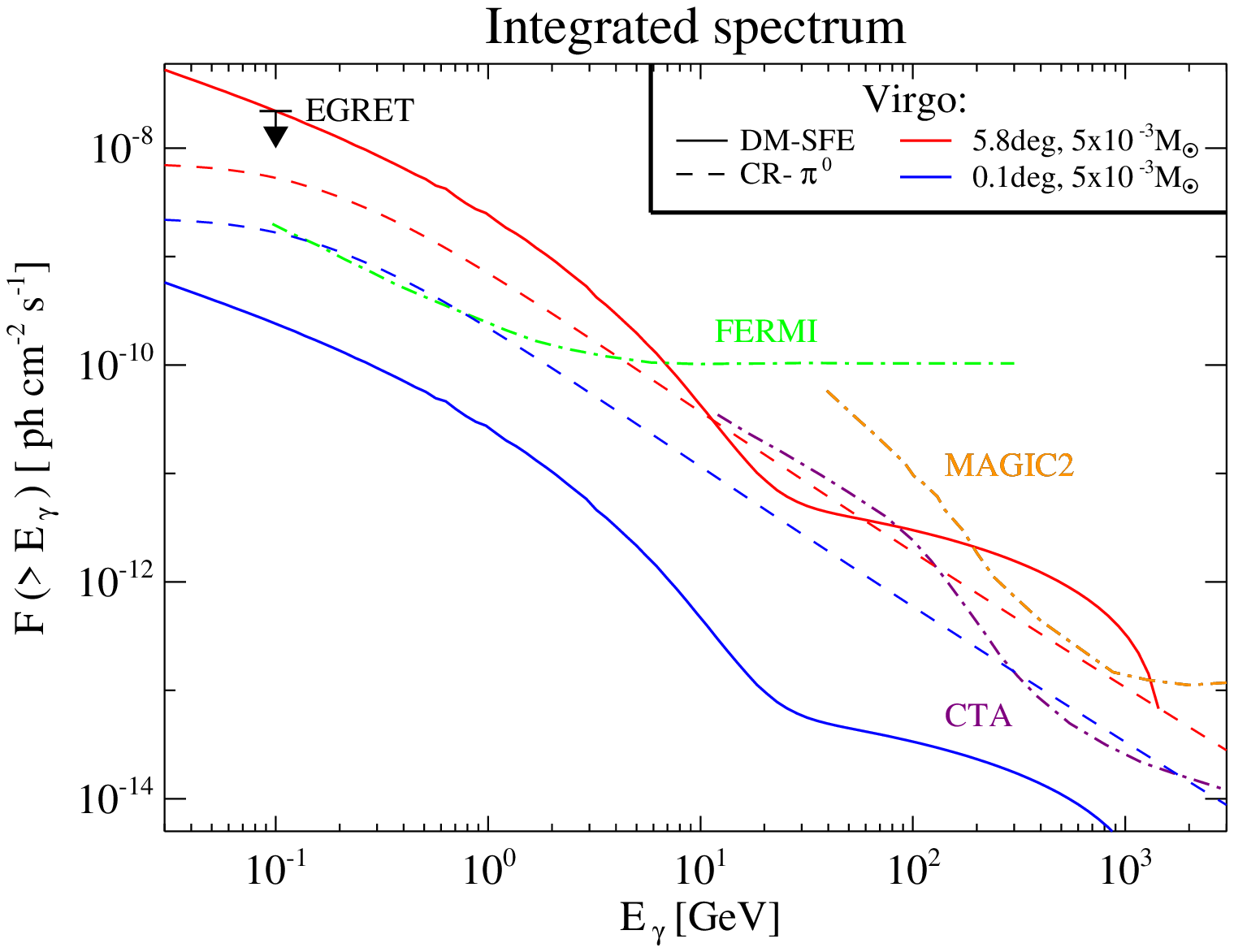}
  \caption{Left: differential spectra for 3 different clusters observed within a
    solid angle of diameter $\Delta \theta = 3.5$ degree (neglecting the
    contribution of the MW's halo). We show the inverse Compton up-scattered CMB
    photons from $e^+$/$e^-$ pairs that themselves result from DM annihilating
    to $\mu^+$/$\mu^-$ (solid) as well as the final state radiation
    (dash-dotted).  We include Sommerfeld enhancement (SFE) and the enhancement
    from cluster substructures down to a limiting substructure mass of
    $M_\rmn{lim} = 5\times10^{-3}\,\rmn{M}_\odot$. The $\gamma$-ray emission
    from decaying neutral pions that result from hadronic cosmic ray (CR)
    interactions with the ambient gas is shown with dashed lines for our
    conservative model without galaxies. Right: we contrast the integrated
    spectrum of Virgo for the EGRET angular resolution, $\Delta \theta = 5.8$
    (red), with that of imaging air \v{C}erenkov telescopes, $\Delta \theta =
    0.1$ (blue), and compare those to the point source sensitivity curves on the
    5$\sigma$ level of Fermi (2 year all-sky survey) as well as MAGIC2 and CTA
    (50 hours). We choose $M_\rmn{lim} = 5\times10^{-3}\,\rmn{M}_\odot$, so that
    the resulting flux is just consistent with the EGRET upper limit
    \cite{Reimer}. This reduces the substructure boost from $\sim$ 220 to 50.}
\label{fig:diff/int}
\end{minipage}
\end{figure*}

\begin{figure*}
\begin{minipage}{2.0\columnwidth}
  \includegraphics[width=0.48\columnwidth]{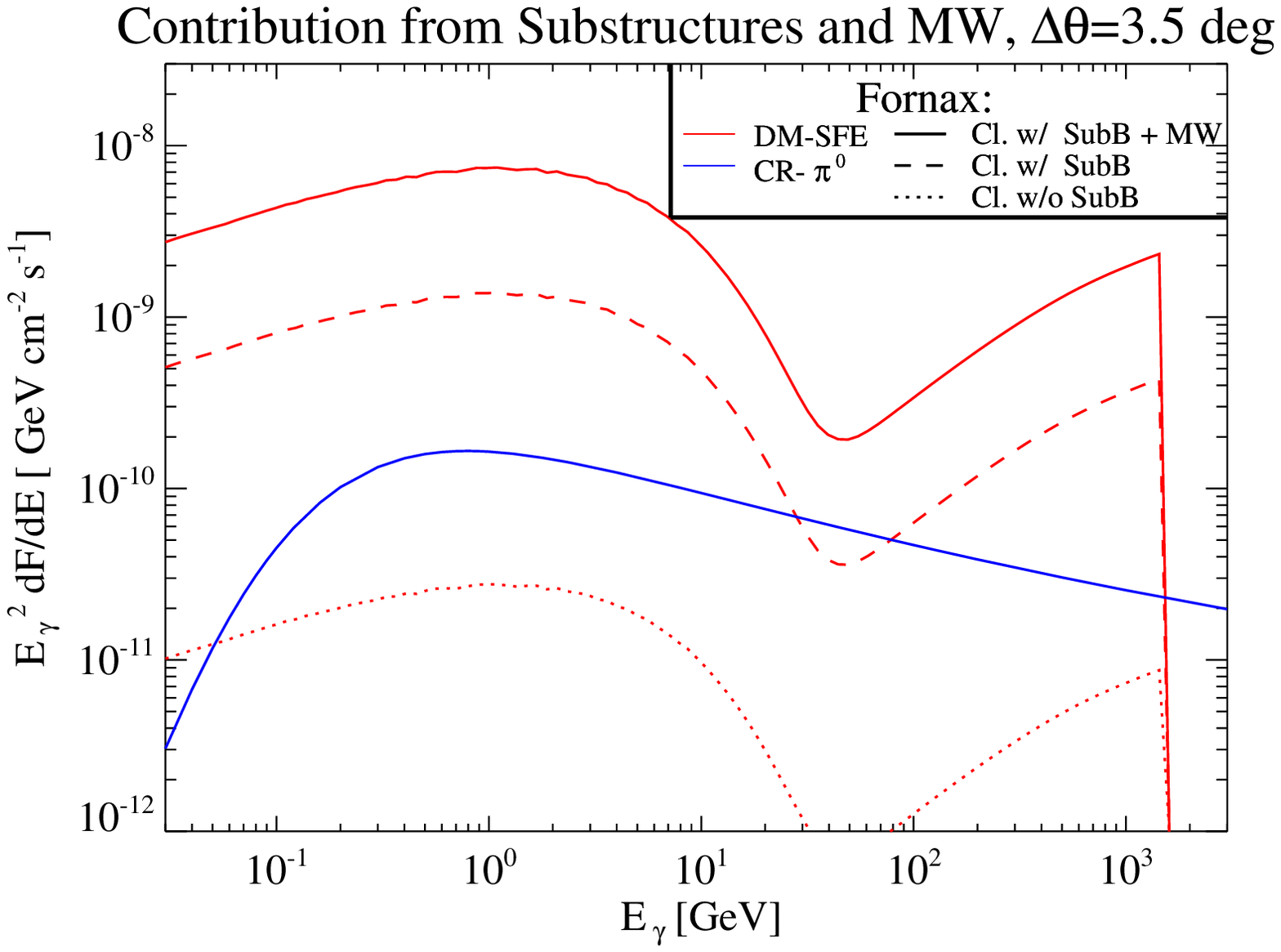}
  \includegraphics[width=0.48\columnwidth]{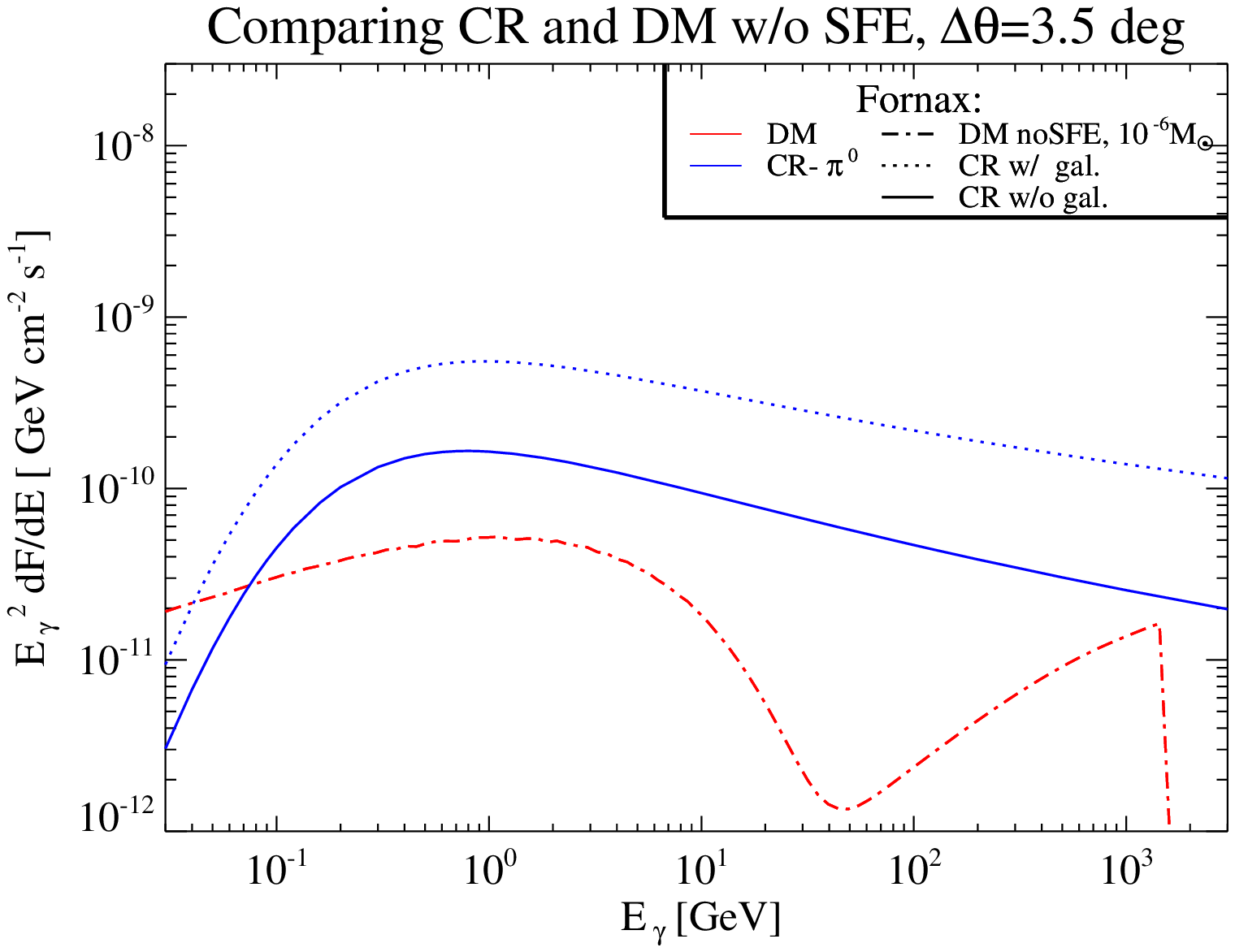}
  \caption{Studying separately the contribution from substructure and the SFE to
    the differential spectra of Fornax within a solid angle of $\Delta \theta =
    3.5$ degree. Left: the dark matter annihilation spectra with SFE and
    $M_\rmn{lim} = 5\times10^{-3}\,\rmn{M}_\odot$ (red) is compared to the pion
    decay spectrum from CR interactions (blue). We show the pure contribution
    from the smooth cluster halo (dotted, $\B_\rmn{sfe}=490,\B_\rmn{sub}=1$), to
    which we add substructures (dashed, $\B_\rmn{sfe}=490,\B_\rmn{sub}=50$), and
    to which we additionally add the line-of sight contribution due to the Milky
    Way's halo towards Fornax (solid). Right: we compare the hadronically
    induced pion decay spectrum to the DM annihilation signal {\em without
      SFE}. The pion decay spectrum shown with (dotted) and without (solid) the
    contribution due to galaxies and dense point sources. The substructure
    boosted DM annihilation signal including the MW contribution (dash-dotted),
    but assuming a standard value for the limiting substructure mass of
    $M_\rmn{lim} = 10^{-6}\,\rmn{M}_\odot$.}
\label{fig:theory}
\end{minipage}
\end{figure*}

{\bf Results and Discussion.}
In Fig.~\ref{fig:diff/int}, we find that, given our assumptions, the DM
annihilation signal in Fornax and Virgo should be clearly visible by Fermi. It
dominates over the CR induced signal for both of our CR models. The annihilation
signal within $3.5\deg$ is similar for Fornax and Virgo -- these are clusters
with a comparable distance but the latter being twice as heavy: the larger
signal of Virgo due to its larger mass is counteracted by the larger
substructure boost of Fornax for the same angular extent.  Using the standard
assumptions for the limiting mass of substructures within DM halos of $10^{-6}\,
\rmn{M}_\odot$, we show that the resulting annihilation flux from Virgo
is already in conflict with the EGRET upper limit. This allows us to place a lower
bound on the limiting mass $M_\rmn{lim} = 5\times10^{-3}\, \rmn{M}_\odot$ and
hence to constrain the free streaming scale in the linear matter power spectrum
to
\begin{equation}
  \label{eq:freestream}
  k < \frac{6\pi}{16} \left(
  \frac{4\pi}{3}\frac{\Omega_\rmn{m}\rho_\rmn{cr}}{M_\rmn{lim}}\right)^{1/3}
  \simeq 35\, \rmn{kpc}^{-1}.
\end{equation}
The Fermi sensitivity will allow us to place an even more stringent limit of
$M_\rmn{lim} > 10^3\,\rmn{M}_\odot$ which is approaching the upper limit
$M_\rmn{lim} < 2 \times 10^{8}\,\rmn{M}_\odot$ derived from Ly-$\alpha$ power
spectrum measurements \cite{wdmconstraints}.  The contribution of the smooth DM
halo component towards high galactic latitudes within 3.5 degree amounts to
$F(>100\,\rmn{MeV}) \simeq 9 \times10^{-9}\,\rmn{ph~cm}^{-2}\,\rmn{s}^{-1}$
which should be easily detectable by Fermi, especially considering an
enhancement factor of a few from substructure in the MW halo.  This allows us to
finally scrutinize the DM models motivated by the recent Fermi, H.E.S.S., and
PAMELA data.  Assuming SFE, the next generation of imaging air \v{C}erenkov
telescopes have good prospects of detecting the DM annihilation signal while it
will be very difficult without an enhancement. We show in Fig.~\ref{fig:theory},
that the DM annihilation flux is substantially boosted due to substructures in
clusters as well as in the MW's halo that has a smooth angular emission
characteristic but is expected to have the same spectral behavior. This provides
hope that even in the absence of SFE, the DM annihilation flux is of the same
order of magnitude as our conservative model of CR induced $\gamma$-ray
emission. The very distinctive spectral properties of the DM-induced
$\gamma$-rays and the universality of the CR spectra suggest that spectral
subtraction techniques could be applied to detect the annihilation signal and
characterize the properties of DM.

{\bf Conclusions.} The DM models motivated by the recent Fermi, H.E.S.S., and
  PAMELA measurements require an anomalous boost factor of 1100. Assuming that
  SFE entierly accounts for this boost, this necessarily predicts large
  annihilation fluxes from nearby galaxy clusters even in the case of somewhat
  reduced SFE due to the larger velocity dispersion of clusters. Using standard
assumptions for the limiting mass of substructures within DM halos, we find a
violation of the EGRET upper limit in Virgo. The lighter a DM particle, the
larger the induced free-streaming scale in the power spectrum and the higher the
mass cutoff for the smallest structures: since the DM interpretation fixes the
DM particle mass this locks in a minimum substructure mass. Hence, a
non-detection of $\gamma$-rays at the predicted level by Fermi would provide a
serious challenge for the standard assumptions of the CDM power spectrum, or
call for a new dynamical effect during non-linear structure formation that wipes
out the smallest structures. The resolution may of course also be that the
rising positron ratio measured by PAMELA and the electron plus positron excess
seen by H.E.S.S. and Fermi is caused by local astrophysical sources,
e.g. pulsars, and is unrelated to dark matter.

{\bf Acknowledgements.}  L.B. and A.P. are grateful to the Swedish Research
Council (VR) for financial support and the Swedish National Allocations
Committee (SNAC) for the resources granted at HPC2N. C.P. thanks P. McDonald for
valuable comments.



\vspace{-0.7cm}
\begin{thebibliography}{00}

\bibitem{pamela_positrons}
O.~Adriani {\it et al.}  [PAMELA Collaboration],
  Nature {\bf 458}, 607 (2009).

\bibitem{atic}
  J.~Chang {\it et al.},
  Nature {\bf 456} (2008) 362.

\bibitem{fermi} The Fermi/LAT collaboration, see e.g.
  W.~B.~Atwood {\it et al.}  [LAT Collaboration],
  Astrophys.\ J.\  {\bf 697} (2009) 1071.

\bibitem{fermicre}
A.A.~Abdo {\it et al.} [Fermi/LAT Collaboration],
Phys.\ Rev.\ Lett.\ {\ bf 102}, 181101 (2009).

\bibitem{hess}
  F.~Aharonian {\it et al.}  [H.E.S.S. Collaboration],
  Phys.\ Rev.\ Lett.\  {\bf 101}, 261104 (2008).

\bibitem{hooper}
  D.~Hooper,
  arXiv:0901.4090 [hep-ph].

\bibitem{lberev}
  L.~Bergstr\"om,
  New J.\ Phys.\  {\bf 11}, 105006 (2009).

\bibitem{pulsars}
  D.~Grasso {\it et al.}  [FERMI-LAT Collaboration],
  Astropart.\ Phys.\  {\bf 32} (2009) 140.

\bibitem{bez}
  L.~Bergstr\"om, J.~Edsj\"o and G.~Zaharijas,
  Phys.\ Rev.\ Lett.\  {\bf 103}, 031103 (2009).

\bibitem{NFW}
  J.~F.~Navarro, C.~S.~Frenk, S.~D.~M.~White,
  Astrophys.\ J.\  {\bf 490}, 493 (1997).

\bibitem{Draco} 
  S. Mashchenko, A. Sills, and H.~M.~Couchman, 
  Astrophys.\ J.\  {\bf 640}, 252  (2006).

\bibitem{subhalo}
L.~Gao, et al., Mon.\ Not.\ R.\ Astron.\ Soc.\ {\bf 355}, 819 (2004).

\bibitem{colafrancesco}
  S.~Colafrancesco, S.~Profumo, and P.~Ullio,
  Astron.\ Astrophys.\  {\bf 455}, 21 (2006).

\bibitem{previous}
  T.~E.~Jeltema, J.~Kehayias and S.~Profumo,
  Phys.\ Rev.\  D {\bf 80} (2009) 023005.

\bibitem{sommerfeld}
  J.~Hisano, S.~Matsumoto, and M.~M.~Nojiri,
  Phys.\ Rev.\ Lett.\  {\bf 92}, 031303 (2004);
   N.~Arkani-Hamed et al., Phys.\ Rev.\ D {\bf 79}, 015014 (2009).

\bibitem{resonance} 
  M.~Kuhlen and D.~Malyshev,
  Phys.\ Rev.\  D {\bf 79}, 123517 (2009).

\bibitem{SFEformula}
  M.~Kamionkowski and S.~Profumo,
  Phys.\ Rev.\ Lett.\  {\bf 101} 261301 (2008).

\bibitem{Voit}
  G.~M.~Voit, Rev.\ Mod.\ Phys.\ {\bf 77}, 207 (2005).

\bibitem{HIFLUGCS}
  T.~H.~Reiprich and H.~B{\"o}hringer, Astrophys.\ J.\  {\bf 567}, 716
  (2002).

\bibitem{maccio_cfit}
  A.~V.~Maccio', A.~A.~Dutton and F.~C.~v.~Bosch,
  arXiv:0805.1926 [astro-ph].

\bibitem{zhao_cfit}
  D.~H.~Zhao et al.,
  arXiv:0811.0828 [astro-ph].

\bibitem{c_conv}
  W.~Hu and A.~V.~Kravtsov,
  Astrophys.\ J.\  {\bf 584} (2003) 702.


\bibitem{M_Virgo} 
S.~Schindler, B.~Binggeli, and H.~B{\"o}hringer,  Astron.\ Astrophys.\  
{\bf 343}, 420 (1999).

\bibitem{Cholis:2008wq}
  I.~Cholis et al.,
  arXiv:0811.3641 [astro-ph].

\bibitem{Einasto}
  J.~F.~Navarro {\it et al.},
  arXiv:0810.1522 [astro-ph].

\bibitem{Aquarius}
  V.~Springel {\it et al.},
  arXiv:0809.0898 [astro-ph].


\bibitem{subboost}
  V.~Springel {\it et al.},
  Nature {\bf 456N7218} 73 (2008).

\bibitem{Mlim}
  S.~Hofmann, D.~J.~Schwarz and H.~Stoecker,
  Phys.\ Rev.\  D {\bf 64} (2001) 083507;
  A.~M.~Green, S.~Hofmann and D.~J.~Schwarz,
  JCAP {\bf 0508} (2005) 003.
  
\bibitem{Springel:2005mi}
  V.~Springel,
  Mon.\ Not.\ Roy.\ Astron.\ Soc.\  {\bf 364} (2005) 1105.

\bibitem{paperII}
  C.~Pfrommer, T.~A.~En{\ss}lin, and V.~Springel, Mon.\ Not.\ R.\ Astron.\ Soc.\ 
  {\bf 385}, 1211 (2008).

\bibitem{TeVpap}
  A.~Pinzke and C.~Pfrommer (to be published).

\bibitem{piondecay} 
C.~Pfrommer and T.A.~En{\ss}lin,  Astron.\ Astrophys.\  {\bf 413}, 17 (2004).

\bibitem{electdens}
  M.~Paolillo et al.,
  Astrophys.\ J.\  {\bf 565} (2002) 883;
  K.~Matsushita et al.,
  arXiv:astro-ph/0201242;
  U.~G.~Briel, J.~P.~Henry, and H.~Boehringer, 
  Astron.\ Astrophys.\  {\bf 259}, L31 (1992).

\bibitem{paperIII}
C.~Pfrommer, Mon.\ Not.\ R.\ Astron.\ Soc.\ {\bf 385}, 1242 (2008).

\bibitem{wdmconstraints}
U.~Seljak et al., Phys. Rev. Lett. {\bf 97}, 191303 (2006)

\bibitem{Reimer} 
O.~Reimer et al.,  Astrophys.\ J.\ {\bf 588}, 155 (2003).


\end{thebibliography}
\end{document}